\documentclass{ws-procs975x65}
\usepackage{graphicx}
\usepackage{mediabb}
\def\beq{\begin{equation}}
\def\eeq{\end{equation}}
\def\etal {{\it et al.}}

\newcommand{\Ref}[1]{Ref.~\refcite{#1}}
\newcommand{\Refs}[1]{Refs.~\refcite{#1}}
\newcommand{\zu}[1]{Fig.~\ref{#1}}

\newcommand{\Msun}{M_{\odot}}
\begin{document}

\title{Prospects for improving the sensitivity of\\ KAGRA gravitational wave detector}
\author{Yuta~Michimura,$^{a*}$ %
Masaki~Ando,$^{a}$ %
Eleonora~Capocasa,$^{b}$ %
Yutaro~Enomoto,$^{a}$ %
Raffaele~Flaminio,$^{b,c}$ %
Sadakazu~Haino,$^{d}$ %
Kazuhiro~Hayama,$^{e}$ %
Eiichi~Hirose,$^{f}$ %
Yousuke~Itoh,$^{g}$ %
Tomoya~Kinugawa,$^{h}$ %
Kentro~Komori,$^{a,i}$ %
Matteo~Leonardi,$^{b}$ %
Norikatsu~Mio,$^{j}$ %
Koji~Nagano,$^{f}$ %
Hiroyuki~Nakano,$^{k}$ %
Atsushi~Nishizawa,$^{l}$ %
Norichika~Sago,$^{m}$ %
Masaru~Shibata,$^{n,o}$ %
Hisaaki~Shinkai,$^{p}$ %
Kentaro~Somiya,$^{q}$ %
Hiroki~Takeda,$^{a}$ %
Takahiro~Tanaka,$^{o,r}$ %
Satoshi~Tanioka,$^{b,s}$ %
Li-Wei~Wei,$^{t}$ %
Kazuhiro~Yamamoto$^{u}$}

\address{$^a$Department of Physics, University of Tokyo, Bunkyo, Tokyo 113-0033, Japan\\
$^b$National Astronomical Observatory of Japan, Mitaka, Tokyo 181-8588, Japan\\
$^c$Laboratoire d'Annecy de Physique des Particules (LAPP), Univ. Grenoble Alpes, Universit{\'e} Savoie Mont Blanc, CNRS/IN2P3, F-74941 Annecy, France\\
$^d$Institute of Physics, Academia Sinica, Nankang, Taipei 11529, Taiwan\\
$^e$Department of Applied Physics, Fukuoka University, Nanakuma, Fukuoka 814-0180, Japan\\
$^f$Institute for Cosmic Ray Research, University of Tokyo, Kashiwa, Chiba 277-8582, Japan\\
$^g$Department of Physics, Osaka City University, Sumiyoshi, Osaka 558-8585, Japan\\
$^h$Department of Astronomy, University of Tokyo, Bunkyo, Tokyo 113-0033, Japan\\
$^i$LIGO Laboratory, Massachusetts Institute of Technology, Cambridge, MA 02139, USA\\
$^j$Institute for Photon Science and Technology, University of Tokyo, Bunkyo, Tokyo 113-8656, Japan\\
$^k$Faculty of Law, Ryukoku University, Fushimi, Kyoto 612-8577, Japan\\
$^l$Research Center for the Early Universe (RESCEU), School of Science, University of Tokyo, Bunkyo, Tokyo 113-0033, Japan\\
$^m$Faculty of Arts and Science, Kyushu University, Nishi, Fukuoka 819-0395, Japan\\
$^n$Max Planck Institute for Gravitational Physics (Albert Einstein Institute), Am Muhlenberg 1, Postdam-Golm 14476, Germany\\
$^o$Center for Gravitational Physics, Yukawa Institute for Theoretical Physics, Kyoto University, Sakyo, Kyoto 606-8502, Japan\\
$^p$Faculty of Information Science and Technology, Osaka Institute of Technology, Hirakata, Osaka 573-0196, Japan\\
$^q$Department of Physics, Tokyo Institute of Technology, Meguro, Tokyo 152-8550, Japan\\
$^r$Department of Physics, Kyoto University, Sakyo, Kyoto 606-8502, Japan\\
$^s$The Graduate University for Advanced Studies (SOKENDAI), Mitaka, Tokyo 181-8588, Japan\\
$^t$Max Planck Institute for Gravitational Physics (Albert Einstein Institute), Callinstra{\ss}e, Hannover 30167, Germany\\
$^u$Department of Physics, University of Toyama, Toyama, Toyama 930-8555, Japan\\
$^*$E-mail: michimura@granite.phys.s.u-tokyo.ac.jp}

\begin{abstract}
KAGRA is a new gravitational wave detector which aims to begin joint observation with Advanced LIGO and Advanced Virgo from late 2019. Here, we present KAGRA's possible upgrade plans to improve the sensitivity in the decade ahead. Unlike other state-of-the-art detectors, KAGRA requires different investigations for the upgrade since it is the only detector which employs cryogenic cooling of the test mass mirrors. In this paper, investigations on the upgrade plans which can be realized by changing the input laser power, increasing the mirror mass, and injecting frequency dependent squeezed vacuum are presented. We show how each upgrade affects to the detector frequency bands and also discuss impacts on gravitational-wave science. We then propose an effective progression of upgrades based on technical feasibility and scientific scenarios.
\end{abstract}


\bodymatter

%%%%%%%%%%%%%%%%% now a standard article style for the most part

\section{Introduction}
The era of gravitational wave astronomy began with the first direct detections of gravitational waves from binary black holes and binary neutron star systems by Advanced LIGO and Advanced Virgo~\cite{GW150914,GW170817}. Improving the sensitivity of these detectors enables more frequent detections and more precise source parameter estimation. To this end, there have been extensive studies to improve the sensitivity beyond the detector's original design sensitivity.

Within LIGO Scientific Collaboration and Virgo Collaboration, there are ongoing effort to upgrade Advanced LIGO and Advanced Virgo detectors to {\it A+}~\cite{MillerAplus} and {\it AdV+}~\cite{JeromeAdVplus}, respectively, by around 2024~\cite{ObservationScenario}. The designed sensitivities of A+ and AdV+ are improved over that of Advanced LIGO and Advanced Virgo by roughly a factor of two. The improvement is in part realized by the coating thermal noise reduction either from the mechanical loss reduction of the coating material or from larger beam size. Also, broadband quantum noise reduction is expected by using a 300-m filter cavity to generate frequency dependent squeezed vacuum~\cite{EricFilterCavity,EleonoraFilterCavity2016}. Twofold broadband sensitivity improvement leads to eightfold increase in the detection rate, and halves the parameter estimation error.

KAGRA is another laser interferometric gravitational wave detector which is being built in Japan~\cite{iKAGRA,bKAGRAphase1} and plans to start observation jointly with Advanced LIGO and Advanced Virgo from late 2019. Compared with the other detectors, KAGRA has two technologically unique features: it is constructed at a seismically quiet underground site, and it uses sapphire mirrors at cryogenic temperatures to reduce thermal noise. Therefore, KAGRA has a unique potential to further improve its sensitivity, and upgrading KAGRA will require different approach compared with the other detectors.

In this paper, we discuss the prospects for the upgrade of the KAGRA detector. We start by describing possible technologies that can be applied for upgrading KAGRA and show that different technologies will improve the sensitivity in different frequency bands. We then discuss impacts on gravitational wave detections for each upgrade, and show possible strategy for the KAGRA upgrade in this decade.

\section{Technologies for the KAGRA upgrade}
\begin{figure}[t]
  \begin{center}
    \includegraphics[width=0.49\hsize]{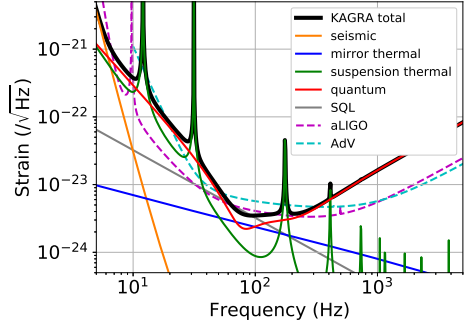}
    \caption{The design sensitivity of KAGRA. The seismic noise shown includes the estimated Newtonian noise from the surface and bulk motion of the mountain containing KAGRA. The mirror thermal noise shown is the sum of the thermal noise from the test mass substrates and the coatings. Sensitivity curves for Advanced LIGO (aLIGO)~\cite{UpdatedaLIGODesign} and Advanced Virgo (AdV)~\cite{ObservationScenario} are also shown for comparison.} \label{bKAGRA}
  \end{center}
\end{figure}

The current design sensitivity of KAGRA is shown in \zu{bKAGRA}. At low frequencies, the sensitivity is limited by the suspension thermal noise and the quantum radiation pressure noise. At high frequencies, the sensitivity is limited by the quantum shot noise. At the most sensitive band in the mid-frequencies, the sensitivity is limited by the mirror thermal noise, which manly comes from the coating Brownian noise. Thanks to cryogenic cooling of the sapphire test masses to 22~K, the mirror thermal noise is smaller than Advanced LIGO and Advanced Virgo although the size of the test mass is smaller. However, the suspension thermal noise is higher since the heat extraction is done by the sapphire fibers suspending the test mass and it requires thick and short fiber (1.6~mm diameter, 35~cm long) for efficient heat extraction. The quantum shot noise is also higher due to input laser power limitation for cryogenic cooling. Because of these features, KAGRA plans to use quantum non-demolition techniques such as the detuing of the signal recycling cavity and homodyne readout to reduce quantum noise in the most sensitive band at the cost of narrowing the detector bandwidth. Detailed discussion on the sensitivity optimization of KAGRA is given in \Refs{SomiyaKAGRA,PSOKAGRA}

To improve the sensitivity of KAGRA, retuning of laser power and suspension parameters will help at certain frequency bands. Increasing the mirror mass and injection of frequency dependent squeezed vacuum are also promising ways to improve the sensitivity. In the following subsections, we will discuss the effect of each technology for the upgrade of KAGRA. We will then discuss longer term prospects for the upgrade which can be realized by combining multiple technologies in this decade. Example sensitivity curves of KAGRA upgraded with different technologies discussed below are shown in \zu{KAGRAplus} (Left). The interferometer parameters and the dimensions of the suspension fibers to calculate these sensitivity curves are optimized with particle swarm optimization method described in \Ref{PSOKAGRA}. The sensitivity curve data are available at \Ref{T1809537}.

\begin{figure}[t]
  \begin{center}
    \begin{tabular}{c}
\begin{minipage}{0.49\hsize}
  \begin{center}
    \includegraphics[width=\hsize]{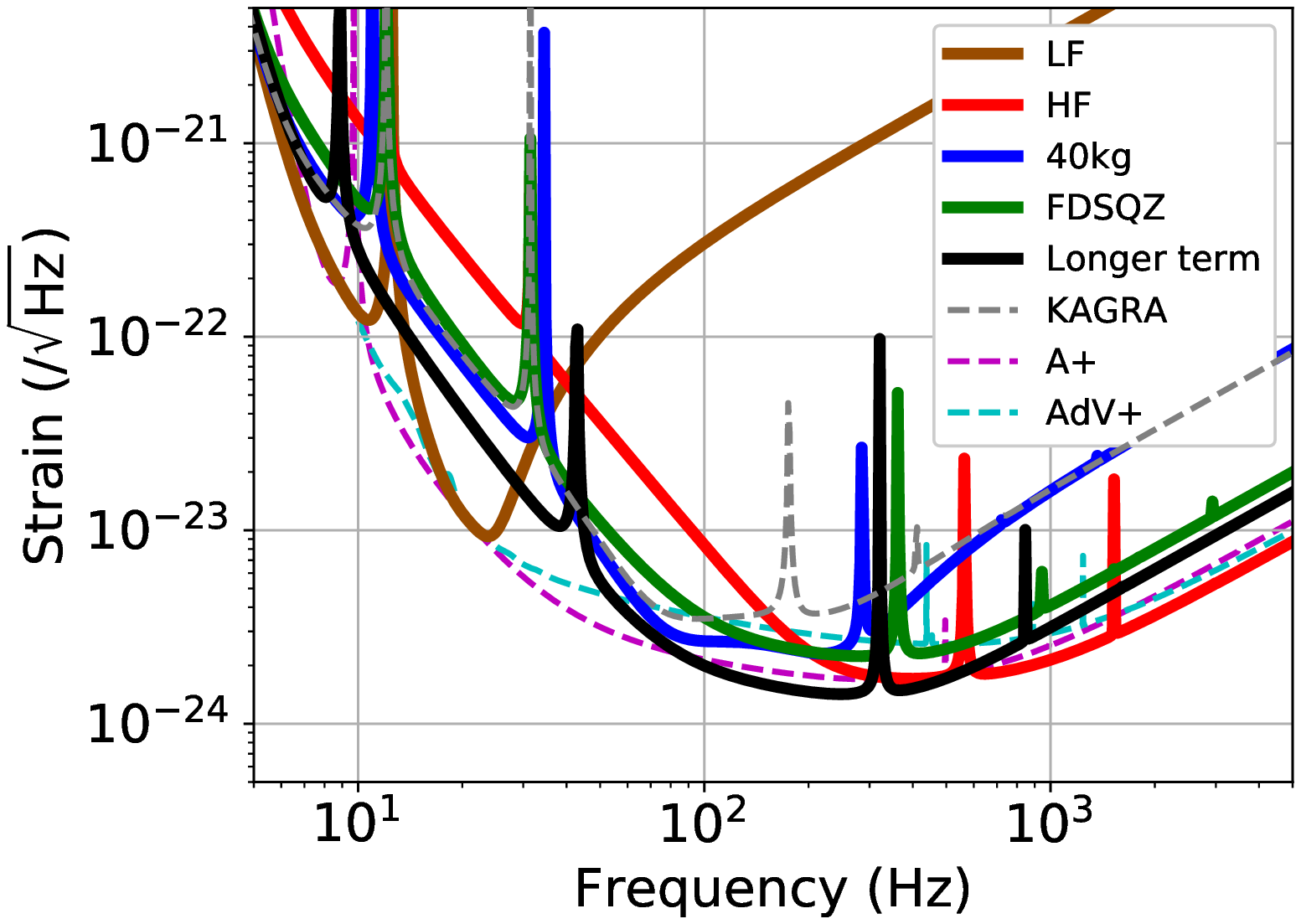}
  \end{center}
\end{minipage}
\begin{minipage}{0.49\hsize}
  \begin{center}
    \includegraphics[width=\hsize]{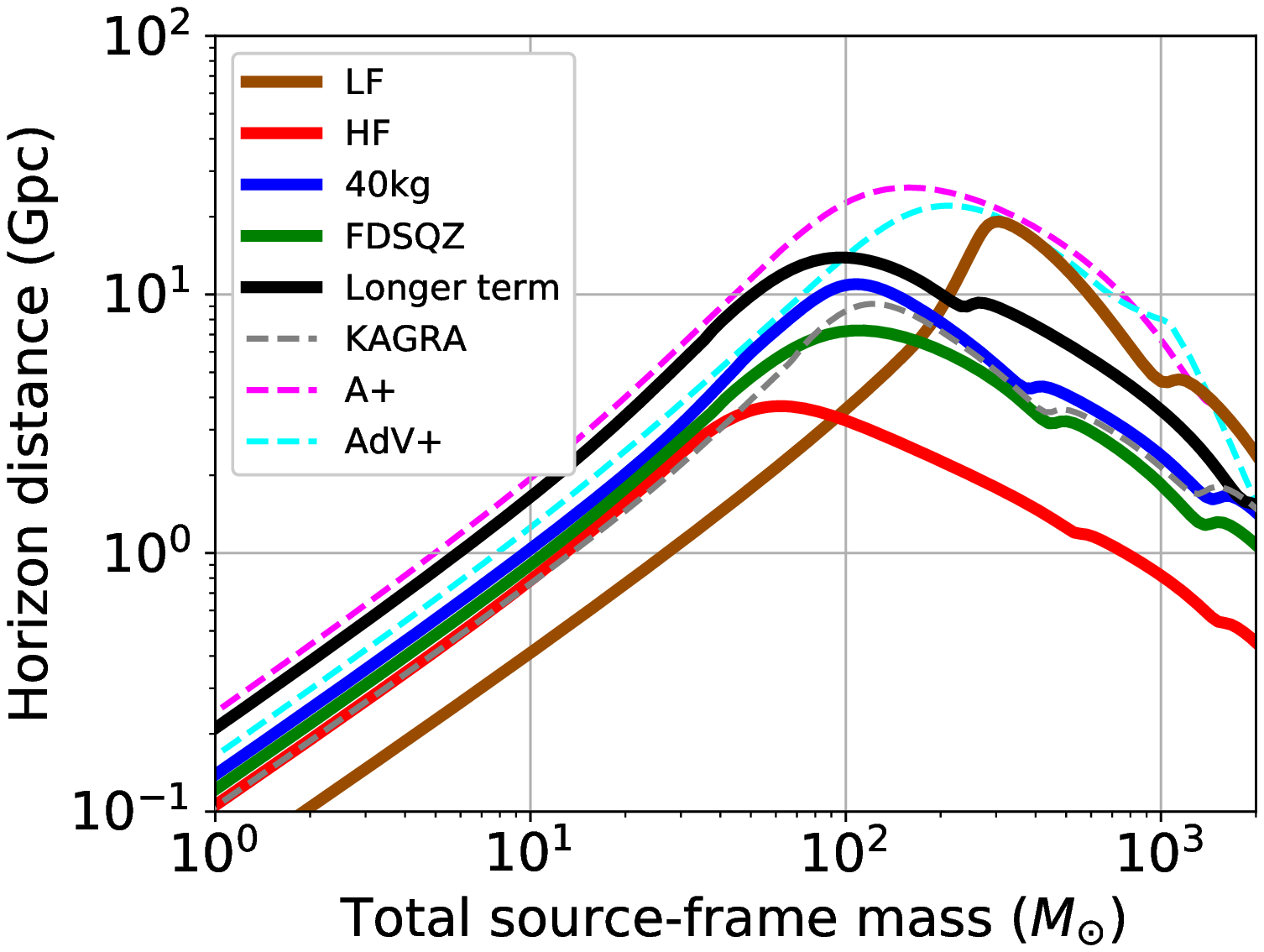}
  \end{center}
\end{minipage}
    \end{tabular}
    \caption{(Left) Example sensitivity curves for the upgrade of KAGRA using different technologies. LF: Lower input power plan to focus on low frequency. HF: Higher power plan with frequency independent squeezed vacuum to focus on high frequency. 40kg: Sensitivity with increased mass of the test masses from 22.8~kg to 40~kg. FDSQZ: Sensitivity with the injection of frequency dependent squeezed vacuum generated with a 30-m filter cavity. Longer term: Example of longer term upgrade plan combining multiple technologies. Sensitivity curves for A+~\cite{AplusDesign} and AdV+~\cite{ObservationScenario} are also shown for comparison. (Right) The horizon distance of example KAGRA upgrades for equal-mass, nonspinning binaries. The horizon distance shows the maximum distance at which gravitational waves can be detected with signal-to-noise ratio of more than 8.} \label{KAGRAplus}
  \end{center}
\end{figure}

\subsection{Laser power and heat extraction}
The input laser power and suspension thermal noise is closely related in KAGRA since heat extraction is done by the suspension fibers. To improve the sensitivity at low frequencies, reduction of suspension thermal noise is necessary. This can be done by changing the suspension fibers to thinner and longer ones since suspension thermal noise scales with $d_{\rm f}^2/l_{\rm f}$, where $d_{\rm f}$ and $l_{\rm f}$ are the diameter and the length of the fiber, respectively. However, this will result in larger shot noise because the heat extraction efficiency will be less and maximum input laser power allowed will be less. Similarly, higher laser power to reduce shot noise at high frequencies require thicker and shorter suspension fibers, which will result in larger suspension thermal noise.

The {\it LF} curve shown in \zu{KAGRAplus} is an example curve which the sensitivity at low frequencies is improved by lowering the laser power at the beam splitter from 673 W to 5 W. This plan requires higher detuning of the signal recycling cavity to reduce quantum noise at around 20-30~Hz. The suspension thermal noise peak at 31~Hz in the original KAGRA design sensitivity comes from the vertical motion of the intermediate mass suspension. Therefore, to remove this peak from the low frequency band, the {\it LF} plan also requires heavier intermediate mass with thinner and longer suspension wires. The interferometer parameters are optimized to maximize the inspiral range of $100~\Msun$-$100~\Msun$ binary in the detector frame.

The {\it HF} curve shown in \zu{KAGRAplus} on the other hand focuses on the high frequencies by increasing the laser power at the beam splitter to 3400~W. It also assumes the injection of frequency independent squeezed vacuum to further reduce the shot noise. Here, 6~dB of detected squeezing at high frequencies is assumed. The interferometer parameters are optimized to minimize the sky localization error of GW170814-like binary neutron stars~\cite{PSOKAGRA}.

\subsection{Increasing the mirror mass}
Increasing the mass of the test mass generally improves the sensitivity since the suspension thermal noise and quantum radiation pressure noise scales with $m^{-3/2}$ and $m^{-1}$, respectively. The coating thermal noise also can be reduced since larger mirror allows larger beam size on the mirror. Assuming both the aspect ratio of the mirror and the ratio of the beam diameter to the mirror diameter to be the same, the coating thermal noise scales with $m^{-1/3}$.

The {\it 40kg} curve shown in \zu{KAGRAplus} is an example sensitivity with the mirror mass increased from 22.8~kg to 40~kg. Considering the design inside the current KAGRA cryostat, 40~kg would be the size limit without changing the cryostat drastically. The interferometer parameters are optimized to maximize the inspiral range of $1.4~\Msun$-$1.4~\Msun$ binary. We note here that coating thermal noise reduction by larger beam size is assumed but smaller mechanical loss of the coating material is not assumed in the sensitivity calculation to show a feasible plan.

Interestingly, increasing the mirror mass result in the sensitivity improvement only at mid-frequencies where coating thermal noise dominates. This is because heavier mass requires higher laser power to keep the frequency $f_{\rm SQL}$ where quantum noise reaches the standard quantum limit to be the same. In case of KAGRA, $f_{\rm SQL}$ should be as high as possible until the quantum noise reaches the coating thermal noise, if we want to maximize the inspiral range. This is because the frequency dependence of the standard quantum limit ($f^{-1}$) is larger than that of the inspiral signal ($f^{-2/3}$). Therefore, the laser power scales with more than $m$. Higher laser power requires thicker suspension fibers and in the end the suspension thermal noise is not much dependent on the mirror mass.

\subsection{Frequency dependent squeezing}
Injection of frequency dependent squeezed vacuum is a promising way to reduce both radiation pressure noise and shot noise, which can be done without increasing the mirror mass or the laser power.  The {\it FDSQZ} curve shown in \zu{KAGRAplus} is an example curve which can be realized with 30-m filter cavity and 5~dB of detected squeezing at high frequencies. 30-m filter cavity can be constructed along the vacuum tubes of the signal recycling cavity. The interferometer parameters are optimized to maximize the inspiral range of $1.4~\Msun$-$1.4~\Msun$ binary.

As discussed previously, the input laser power should be increased for higher $f_{\rm SQL}$ and this result in slightly worse suspension thermal noise. Also, injection of squeezed vacuum prefers no detuning of the signal recycling cavity. Therefore, injection of frequency dependent squeezed vacuum result in a sensitivity improvement at high frequencies.

\subsection{Longer term prospects}
As we have shown, applying only one of these technologies give sensitivity improvement at certain frequency bands. Combination of multiple technologies is necessary for broadband sensitivity improvement. The {\it Longer term} curve shown in \zu{KAGRAplus} is an example sensitivity for 5 to 10-year upgrade plan which can be realized with 100~kg mirrors, 30-m filter cavity and 3500~W of the laser power at the beam splitter. The interferometer parameters are optimized to maximize the inspiral range of $1.4~\Msun$-$1.4~\Msun$ binary.

The situation is similar to {\it FDSQZ} plan, but because of larger test mass, suspension thermal noise and coating thermal noise are also reduced. In total, twofold broadband sensitivity improvement will be realized.

\section{Science case study and disussion on strategic upgrade}
Although combination of multiple upgrade components is necessary for the broadband sensitivity improvement, upgrades to the detector should be done in an incremental way. Which to be implemented at earlier stages depend on the technological feasibility and impact on gravitational-wave science.

Figure~\ref{KAGRAplus} (Right) shows the horizon distance of each example upgrade for compact binary coalescences. {\it LF} plan has the largest horizon distance above $\sim 200\Msun$ in total mass, whereas {\it 40kg} plan has the largest horizon distance for smaller masses. We can say that {\it LF} has the highest probability of detecting the intermediate mass black holes (IMBHs).

Although the horizon distance is not great, {\it HF} plan gives the smallest sky localization error for binary neutron stars. The median of the sky localization error for GW170817-like binaries calculated with the same method described in \Ref{PSOKAGRA} for {\it LF}, {\it HF}, {\it 40kg} and {\it FDSQZ} are 0.507~deg$^2$, 0.105~deg$^2$, 0.156~deg$^2$ and 0.119~deg$^2$, respectively. For the sky localization of $30~\Msun$-$30~\Msun$ binary black holes, {\it 40kg} gives the smallest error. For constraining neutron star equation of state and for search for continuous waves from pulsars, {\it HF} and {\it FDSQZ} will be the best choices since the sensitivity from 500~Hz to 4~kHz is important for these studies. For the test of general relativity through inspiral-merger-ringdown waveform, broadband configuration such as {\it FDSQZ} and {\it 40kg} would be preferred.

From the technical feasibility point of view, {\it LF} has the largest uncertainty since there are many kinds of low frequency excess noises other than the fundamental noises discussed above, such as scattered light noise, vibration noise from cryocoolers, interferometer controls noise etc. 40~kg test mass would be feasible in next few years, but even larger mirror is required for longer term upgrade. Considering that higher power laser source and squeezed vacuum source are required also for longer term upgrade, implementing these as a first step to focus on high frequency sensitivity improvement would be a strategy for the upgrade. {\it HF} plan is also attractive in that it might be able to do original science because {\it HF} has better sensitivity at high frequencies than A+ and AdV+.

\section{Summary}
Upgrading KAGRA requires simultaneous tuning of the parameters related to thermal noise and those related to quantum noise since the heat extraction is done through the fibers suspending the test mass mirrors. We showed that shifting the detector frequency band of KAGRA is possible by changing the input laser power. We also showed that increasing the mirror mass and injection of frequency dependent squeezed vacuum will improve the sensitivity at mid-frequencies and high frequencies, respectively. Considering the technical feasibility and impact on the detection of gravitational waves, possible strategy for upgrading KAGRA would be to focus on high frequency improvement with higher laser power and squeezed vacuum injection for near term. In a longer term, broadband twofold improvement with frequency dependent squeezed vacuum injection and heavier mirror would be realized.

In this study, improvements in the coating, increased heat conductivity of the suspension sapphire fibers and reduced heat absorption of the sapphire mirror are not considered. More detailed investigations and other possibilities of the upgrade will be reported elsewhere.

\end{document}